# Crystal Irradiation Stimulation of Enzyme Reactivity: An Explanation

George E. Bass

Department of Pharmaceutical Sciences, University of Tennessee Health Science Center, Memphis, TN 38163

Abstract

In 1968, S. Comorosan first reported a phenomenon wherein irradiation of the substrate of an enzyme reaction, in the crystalline state, for a specific number of seconds could lead to an enhanced aqueous solution reaction rate (up to 30%). Moreover, for a given reaction, a set of such activating irradiation times, t*, could be found which obey the relationship $t^* = t_m + n\,\tau$ where $t_m$ is the shortest activating time, n = 0, 1, 2, …, and $\tau$ is the constant number of seconds separating consecutive activating times. All $t_m$ and $\tau$ were found to be small multiples of 5 seconds. The basis for this unusual phenomenon has remained a mystery. Previously unreported experimental results are presented which demonstrate, for the lactic dehydrogenase / pyruvate reaction, that the identity of the crystalline material irradiated is, largely, inconsequential. It is proposed here that the irradiation procedure drives oscillatory reactions involving atmospheric gases adsorbed on the crystals. This feature of the model accounts for the wide range of crystalline materials with which the phenomenon has been observed and the peculiar irradiation time dependence. Additional experimental results reveal that lactic dehydrogenase isolated from chicken heart responds to a different set of irradiation times than that isolated from mammalian sources. This behavior, along with published studies by Comorosan, leads to the conclusion that multiple gas-derived bio-active chemical entities are produced, some simultaneously, by the crystal irradiation procedure. Thus, the model derived here posits a set of small molecules which, somewhat analogous with the nitric oxide – guanylate cyclase interaction, are capable of stimulating reaction rates of a wide variety of enzymes and may portend existence of an extended cellular signaling system.

## Introduction

- Description of the Phenomenon

In the late 1960s, Sorin Comorosan (1968), strongly influenced by the theoretical work of Nicholas Rashevsky (1960) and Robert Rosen (1961), set out to develop an experimental approach to the study of quantum theoretic eigenstates and eigenvalues for presumed "biological observables." As a point of departure, he reasoned that biological molecules possess quantum states that can be revealed (observed) only by using highly sensitive biological instruments. He selected enzymes as the instrument and sought to perturb and observe excited states of the molecules which served as their substrates. He would do this by irradiating the pure crystalline substrates and measuring the effect thereof on the subsequent enzyme reaction rate. The phenomenon he and his coworkers uncovered will be referred to here with the acronym CISER (crystal irradiation stimulation of enzyme reactivity).

The basic observation is that the enzyme reaction rate will be increased (typically 10% – 30%) for only certain irradiation times, designated as t*. Other irradiation times have no affect on the reaction rate. For example, if t* happens to be 5.0 sec, a crystalline sample (e.g., sodium pyruvate) irradiated for 2.0, 3.0, 4.5, 5.5 or 7.0 sec would be found to have the same enzyme reaction rate (eg., with lactic dehydrogenase, LDH) as a non-irradiated (t = 0.0 sec) sample. Indeed, it has been found that for a given enzyme reaction the set of all t* values, those irradiation times which produce an increased reaction rate, can be represented as: $t^* = t_m + n\,\tau$, where $t_m$ is the shortest irradiation time in the set (eg., 5.0 sec), n = an integer (0, 1, 2, 3, …) and $\tau$ is a constant time period (e.g., 30.0 sec) which separates successive activating irradiation times (see Fig. 1). Values found for $t_m$ and $\tau$ have always been a multiple of 5.0 seconds (Comorosan, 1980a). The time-width of the activating irradiations is less than ±0.5 sec (Comorosan, 1974; Bass, 1976). The initial studies were conducted using x-rays. Later, it was found that high pressure mercury lamps and some tungsten lamps could be used. Narrow wavelength bands selected with band-pass optical filters covering the visible spectrum produced the phenomenon, most effectively at 546 nm (green) (Comorosan, et al., 1971c, 1972b). The intensity threshold appears to be at an illuminance of 200 – 300 footcandles for a wavelength of 546 nm (Comorosan, 1976, p.195). Values for the $\{t_m , \tau\}$ pair of parameters can vary from one enzyme to the next and sometimes for the same enzyme isolated from different species (e.g., microbial vs. mammalian) (Comorosan,1975b). Patterns in the t* values for different enzyme reactions raise the possibility that an underlying metabolic messaging capability is being reflected (Comorosan, 1971a, 1971b). This behavior has been found also for the effect of irradiated growth factors and antibiotics on growth rate of microorganisms in minimal media and broth, respectively (Comorosan, 1970c, 1975a, 1976; Bass, 1973; Sherman, 1974).

- Scope

A total of 24 enzymes have been investigated and found to respond to the CISER photo-activation process. For many of these, both the forward and reverse reactions have been studied. Some enzymes were obtained from commercial sources, others were isolated by the investigators. Enzyme sources include bacteria, yeast, jack bean, rat, rabbit, pig, beef, human, and chicken. Including the microbial growth rate studies, a total of 34 crystalline compounds (enzyme substrates, cofactors, growth factors, antibiotics, and inorganic salts) have been utilized. These, including experimental studies reported here, are identified in Table 1.

- Validity

Successful CISER experiments have been conducted in at least 6 geographically dispersed laboratories under independent leadership (Bucharest, Romania / Comorosan (1968); Detroit, US / Sherman (1973); Sussex, UK / Goodwin (1975); Memphis, US / Bass (1973); Kansas City, US / Grisolia (1975); and Blacksburg, US / Etzler (1986)). Included in these is a rigorous double-blind study of the urea / urease reaction (Bass *et al.*, 1976a). The latter study established that photo-activation occurred for $t_m$ = 25.0 sec (relative to control values only for irradiation times of 0, 20.0, 24.0, 24.5, 25.5, 26.0, and 30 sec) as well as at 50.0 sec (relative to irradiation times of 0, 50.0 and 60.0 sec) and at 85.0 sec (relative to irradiation times of 0, 80.0 and 90.0 sec). A single-blind confirmation of photo-activation for the LDH / pyruvate reaction for irradiation times of 5 sec (relative to 0, 4 and 6 sec irradiations) and 35 sec (relative to 0, 34 and 36 sec irradiations) has also been reported (Bass & Chenevey, 1976b).

- Previous Explanations

In the first report of experiments involving this phenomenon, Comorosan's attention was focused on the enzyme substrate as the entity whose quantized "biological observables" would be perturbed by irradiation and revealed by subsequent enzyme reaction rates (Comorosan, 1968). Thus, it was the purified crystalline substrates (urea, sodium isocitrate, potassium malate and sodium glutamate) which were irradiated prior to dissolution and measurement of their enzyme reaction rates (urease, isocitrate dehydrogenase, malic dehydrogenase and glutamate dehydrogenase, respectively). However, in two following studies (Comorosan *et al.*, 1970a, 1970c), he found that a non-reactant (cytidine or thymine) could be irradiated for the same t* times, mixed with the crystalline substrate (glucose-6-phosphate or xanthine, respectively) and achieve the same enhanced enzyme reaction rate (glucose-6-phosphate dehydrogenase or xanthine oxidase, respectively). His interpretation at the time was that the irradiated non-reactant crystals transferred the perturbation to the substrate crystals. This facet of the phenomenon was not touched on again until a report in 1980 which detailed experiments wherein irradiated crystalline sodium chloride was effective in stimulating the glutamate-pyruvate transaminase (GPT) conversion of alanine to pyruvate (Comorosan *et al.*, 1980b). Comorosan suggested that the irradiated crystals transfer the perturbation to water molecules which then interact with the enzyme such as to produce a higher reaction rate. In 1988, Comorosan suggested an exciton based model wherein a “phonon wind” would modify the energetic topological configuration of the crystal lattice in a manner consistent with any type of crystal (Comorosan, 1988). Through this period, a number of mathematical modeling studies were reported by Comorosan and co-workers (Comorosan *et al.*, 1975a, 1975b, 1980a).

Previous attempts to explain the CISER phenomenon do little to point the way to verification nor provide a path for subsequent integration of the phenomenon into scientific knowledge. Here, based on previously unreported experimental findings using irradiation of non-reactant crystalline samples, an explanation of the CISER phenomenon will be proposed which attributes the enzyme reaction rate enhancements to photoproducts of atmospheric gases adsorbed on the crystalline samples. Additional experimental results obtained with LDH from different species lead to the proposal that multiple bio-active entities are formed in surface catalyzed free radical mediated oscillatory reactions during the irradiation procedure.

## Materials and Methods

Experiments reported here involve initial reaction rate measurements for lactic dehydrogenase (LDH) conversion of pyruvate to lactate. The reaction rate is represented by the change in absorbance at 340 nm for the associated conversion of NADH to NAD during the first 12 seconds of the reaction, designated $\Delta \mathbf{A_{340}}$. Reaction rates were measured for sets of 4 crystalline samples wherein the first and fourth served as controls (non-irradiated or irradiated for a non-t* time). Each such set is designated a Run. Each Run, from irradiations through reaction rate measurements, was completed within one hour.

All chemicals and biochemicals were obtained from Sigma Chemical Co. Experimental details are provided in Appendix A.

**Results**

- Enzyme Reaction Rate Enhancement by Irradiated Non-reactants

Presented in Table 2 are initial reaction rates (expressed as $\Delta A_{340}$ / 12-sec enzyme reaction) for rabbit muscle LDH conversion of pyruvate to lactate in the presence of solutions of irradiated and control crystalline sodium chloride samples. Enhanced initial reactions rates were found for irradiation times of 5, 35, 155 and 995 sec. These correspond to t* = 5 + 30 n, where n = 0, 1, 5 and 33 (or, $\{t_m,\tau\} = \{5, 30\}$), the same as has been reported previously for irradiated crystalline sodium pyruvate (Comorosan, 1971b, 1972b; Bass, 1976b, 1977). Samples irradiated for t*± 1 sec, as well as samples irradiated 10, 15, 20, 25, and 30 sec, produced no rate enhancement relative to non-irradiated controls. Presented in Tables 3 – 6 are similar results demonstrating rate enhancements for irradiated crystalline potassium chloride, sodium bromide and diatomaceous earth (insoluble, ~ 95% silicon dioxide) for 5 and 35 sec.

- Chicken vs Mammalian LDH

Experiments conducted using irradiated sodium pyruvate with LDH derived from beef heart, porcine heart, and human erythrocytes produced enhanced initial reaction rates for $t^* = 5 + 30\,n$ sec (or, $\{t_m,\tau\} = \{5, 30\}$), just as for rabbit muscle LDH. However, corresponding experiments (irradiated crystalline sodium pyruvate) utilizing LDH isolated from chicken heart yielded a different set of stimulating irradiation times, in particular, $t^* = 15 + 20\,n$ sec (or, $\{t_m,\tau\} = \{15, 20\}$). To confirm that these differences are associated with the enzymes *per se* rather than some aspect of experimental procedure, a series of direct comparison experiments were conducted wherein both enzymes were used in each Run involving 4 identical crystalline samples, 2 irradiated and 2 not. For each irradiation time, two Runs were conducted with opposite sequenceing in the use of the two enzymes. Results are presented in Table 7. These results demonstrate stimulation of the rabbit muscle / pyruvate reaction for irradiation times of 5 and 35 sec, while the chicken heart LDH / pyruvate reaction is stimulated for irradiation times of 15, 35 and 55 sec.

**Discussion**

This report primarily is concerned with the immediate consequence of the irradiation process. That is, what is the activated entity or set of entities, the photoproduct(s), responsible subsequently for increasing enzyme reaction rates? The results presented here for irradiated NaCl, NaBr, KCl and diatomaceous earth serve to verify and extend Comorosan's observations that the phenomenon can be produced by irradiation of a crystalline material that is not involved in the enzyme reaction. Considering the chemical structures and physical characteristics of these compounds, as well as the others presented in Table 1, one can only conclude that the identity of the irradiated crystalline material is of only secondary importance. That is, photoactivation does not rely on the absorption spectrum of the irradiated crystalline material. As shown with diatomaceous earth, it is not even necessary for the crystalline material to be water soluble. For example, generalizing and applying these observations, the reactions of each of the 27 enzymes listed in Table 1 could be stimulated as well by diatomaceous earth irradiated for the various times as found using their irradiated substrates.

There is a notable common circumstance applicable to all studies reported to date. All irradiations have been conducted in the open atmosphere of the laboratory. It is proposed here that the crystal irradiation process induces oscillatory free radical mediated reactions involving atmospheric gases adsorbed on the crystalline materials and are the same whether the crystalline material happens to be the enzyme substrate or some non-reactant. On cessation of irradiation, defining the t* period, much slower dark reactions would lead to relatively stable chemical species which, in turn, are capable of altering reactivity of a particular enzyme.

At this time, there is no direct evidence linking the CISER phenomenon to atmospheric gases. Of course, the essential biological roles played by dissolved gases as reactants and products in respiration and photosynthesis have been know for a long time. However, beginning with discovery in the 1980s of the involvement of nitric oxide in guanylate cyclase stimulation, we have begun to recognize a more subtle role of small gaseous molecules as biological messengers (Stamler et al., 1992; Snyder & Bredt, 1992).

Stimulation of guanylate cyclase by nitric oxide increases $V_{max}$ for conversion of GTP to cGMP by 40- to 50-fold (Wolin, et al., 1982). The possibility of similar signaling functions for carbon monoxide, peroxide and ozone are matters of current investigation (Maines, 1996; Dioum et al., 2002; Nathan, 2002; Wentworth, et al., 2002; Boehning & Snyder, 2002). A corresponding signaling role for CISER photoproducts posited here would be consistent with observations (Comorosan *et al.*, 1971a, 1971b).

Perhaps the most eye-catching feature of the CISER phenomenon is its oscillatory dependence on irradiation time - seemingly strange behavior. However, for a number of reactions involving atmospheric gases very similar oscillatory reaction kinetics, also with fixed periods of seconds and tens of seconds, are well know and characterized. Most extensively studied are the oxidation reactions of hydrogen, carbon monoxide and acetaldehyde (Gray & Scott, 1990). Carbon monoxide oxidation is particularly interesting here.

Mixtures of carbon monoxide and oxygen, in the presence of at least trace amounts of a hydrogen source such as $H_2$ or $H_2O$, can exhibit oscillatory reaction kinetics accompanied by sharp chemiluminescence flashes. For a flow system, reported studies illustrate flashes every 4 sec sustained indefinitely; for a closed system, every 100 seconds for several hours duration (see Fig. 2). The chemiluminescence is referred to as a "light house" effect. It is associated with formation of carbon dioxide via the reaction:

$$CO + O + M \rightarrow CO_2^* : M \rightarrow CO_2 + M + h\nu$$

where M represents a non-reacting third body required for stabilization, $CO_2^*$ is electronically excited, and hν is the photon released when $CO_2^*$ falls to the ground state. The chemiluminescence consists of a continuous spectrum in the wavelength range of 300 to over 500 nm, usually seen as a light blue (Clyne & Thrush, 1962). The chemical entities thought to be kinetically involved (reactants, intermediates and products) via over 33 individual reactions include: CO, $CO_2$, $H_2$, H , $O_2$ , O , $H_2O$, OH, $O_2H$, and $H_2O_2$ . Oscillatory kinetics with a period on the order of 10 sec has also been demonstrated for the gas phase reaction between acetaldehyde and oxygen (see Fig. 3). Chemical entities involved, in addition to all of those above, are thought to include: HCO, $CH_3O$, $CH_2O$, $CH_3CO$, $CH_3CHO$, $CH_2O$, $CH_3OH$, $CH_3$ , $CH_4$ , $C_2H_6$ , $CH_3O_2$, $CH_3CO_2$, $CH_3CO_3$ and $CH_3CO_3H$ (Gray & Scott, 1990, p. 429).

Surface catalyzed oscillatory gaseous reactions have also been observed for carbon monoxide oxidation on platinum and palladium (Scott & Watts, 1981; Mukesh, et al., 1983; Oken & Wicke, 1986; Schwanker et al., 1987; Cirak, et al., 2003). The metal may be present as a bare wire, single crystals, or supported on alumina or xeolytes. Periods are found to be similar to those of the strictly gas-phase reactions (see Fig. 4). Thus, oscillatory, free-radical mediated reactions involving some of the atmospheric gases occur over an appreciable range of conditions. Characteristically, such oscillating systems include a relatively slow initiation reaction that generates the first free radicals as well as propagation, branching and terminating reactions (Gray & Scott, 1990, p. 410). In the studies noted here, reactions were driven by elevated temperatures (700 – 800 K).

Prolonged UV photolysis of carbon monoxide gas in the presence of liquid water has been reported to produce HCHO, $CH_3CHO$, $CH_3OH$, $CH_3COOH$, HCOOH and $CH_3CH_2OH$ (Bar-Nun & Hartman, 1978). Hubbard *et al.* (1971) have reported UV photocatalytic production of organic compounds from carbon monoxide in the presence of moisture and either soil or finely powdered vycor glass. Though an exhaustive identification of photoproducts was not performed, evidence was provided for the presence of glycolic acid, formaldehyde and acetaldehyde. Thus, photolysis of carbon monoxide can produce compounds such as acetaldehyde which, as noted above, can participate in oscillatory reactions. Accordingly, small organic compounds may be considered among the candidates for CISER bio-active photoproducts.

The considerations above lead to the more specific proposal that CISER involves oscillatory free radical mediated reactions on the surface of the irradiated crystalline material. The role of the crystalline material (absorbant) might be simply to concentrate the gas(es), stabilize reactive intermediates, alter polarizabilities such as to increase photo transition probabilities, and/or concentrate photoproducts and follow-on dark reaction species. The irradiation time dependence requires that one or more key reactions are photo-driven. One candidate for this role is the reverse of the CO chemiluminescence reaction, i.e., $CO_2 + M + h\nu \rightarrow CO_2^*{:}M \rightarrow CO + O + M$, where M is a non-reacting third body, possibly the crystal surface. It may be noted that nitric oxide also can engage in a similar chemiluminescent reaction: $O + NO + M \rightarrow NO_2^*{:}M \rightarrow NO_2 + M + h\nu$ (Clyne & Thrush, 1962). While the non-urban area tropospheric concentrations of nitric oxide and nitrogen dioxide are reported as 0 – 300 parts per trillion (Hobbs, 2000), nitric oxide levels probably are higher in the laboratory environment. Humans have been found to exhale nitric oxide at 1 – 200 parts per billion (Byrnes, et al., 1996).

How many different bio-active species are generated during the irradiation process? To get a handle on this question, one can start by examining the irradiation times for the pyruvate reaction of LDH derived from rabbit muscle, chicken heart and yeast. First, note that the rabbit muscle LDH / pyruvate reaction will be stimulated by irradiation times of t* = 5 sec, 35 sec, 65 sec, … . Let us make the simplifying assumption here that only a single recurring photoproduct species is responsible for rate enhancement corresponding to all of the t* irradiation times for a particular enzyme reaction. Designate the photoproduct responsible for the rabbit muscle LDH / pyruvate reaction $\varepsilon_1$. Note however that the chicken heart LDH / pyruvate reaction is not stimulated by a 5 sec irradiation (i.e., $\varepsilon_1$). Instead, the latter is stimulated by an irradiation time of 15 sec. The corresponding photoproduct, call it $\varepsilon_2$, has no effect on the rabbit muscle reaction. Thus, there are at least 2 distinct bio-active entities. One may further note that both the rabbit muscle and chicken heart enzymes are stimulated by an irradiation of 35 sec. With our initial assumption of a single photoproduct being responsible for all stimulations of a given enzyme's reaction, this suggests that $\varepsilon_1$ and $\varepsilon_2$ are present simultaneously at 35 sec of irradiation time. Continuing, the yeast LDH / pyruvate reaction is stimulated by an irradiation time of 25 sec (Comorosan,1975b). This would correspond to still another distinct photoproduct, say $\varepsilon_3$. Since the yeast enzyme also responds to 5 sec irradiations, as does the rabbit muscle enzyme, one may argue that $\varepsilon_1$ and $\varepsilon_3$ are present simultaneously at 5 sec. A more systematic way to make these arguments is to note that, given the initial assumption, a different bio-active entity is indicated for each distinct $\{t_m,\tau\}$ parameter pair, here, {5,30} for rabbit muscle enzyme, {15,20} for chicken heart enzyme, and {5,20} for yeast enzyme.

Next, one may compare the $\{t_m,\tau\}$ parameter pairs for the forward and reverse reactions of a given enzyme. A listing of enzymes for which both forward and reverse reactions have been reported is provided in Table 8. Here, one may observe that for the forward – reverse reaction set of a given enzyme from a given source, the $t_m$ parameters have different values while the $\tau$ parameters are the same. Accordingly, no irradiation time that stimulates the forward reaction has any effect on the corresponding reverse reaction, and conversely. Thus, one is led to propose that different photoproduct entities must be evoked for each occurrence of such distinct $\{t_m,\tau\}$ parameter pairs. Moreover, the extended generalization of this argument leads to the proposal that there exists a different bio-active photoproduct (or subsequent dark reaction species) for every observed $\{t_m,\tau\}$ parameter pair, regardless of the enzyme reaction involved. A matrix of all reported $\{t_m,\tau\}$ parameter pairs, 14 in number, is presented in Table 9. When contemplating this aspect of the model, one must bear in mind that imposition of the assumption made here that only a single photoproduct is responsible for all stimulations of a given enzyme reaction is justifiable solely on the basis of simplification. Further, the studies from which Table 9 is derived were conducted over an extended period of time and cover a very wide range of experimental conditions. Degeneracies may exist in the relationship between this set of parameter pairs and a set of bio-active photoproducts.

The model proposed here can be summarized as follows. Irradiation of crystalline material in the open atmosphere with visible light can drive oscillatory free radical mediated reactions involving adsorbed gas molecules. These reactions generate multiple transitory photoproducts and/or follow-on dark reaction species, a subset of which are capable of stimulating reaction rates of a wide range of enzymes.

The biological significance of the CISER phenomenon remains undetermined. Patterns noted earlier in the $\{t_m,\tau\}$ parameters for metabolic enzymes isolated from yeast and rabbit muscle, as well as manifestation in growth rates of several microbial species, are suggestive of a possible underlying signaling system. How the stimulating species might be generated *in vivo* remains an open question. But, this is not unlike early questions concerning nitric oxide and carbon monoxide.

As is illustrated in Table 1, studies of the CISER phenomenon have encompassed 24 enzymes and 12 living species. One must at least consider the possibility that the phenomenon reflects a rather common characteristic of enzyme catalysis. Might photoproducts similar to those postulated here have been generated by natural processes in the sands, clays and waters to become participants in early evolution and thereby influence emerging enzyme architecture?

This model is verifiable. From an enzyme reaction kinetics perspective, one can argue that the stimulating photoproducts will be present in solution in molar quantities at least as great as the enzyme. For the LDH reactions reported here, this falls in the picomole range. This level of small gaseous molecules can be identified by GC-MS methods (e.g., see Leffler et al., 2003). Once such species are identified, they can be generated by other means, possibly more efficiently, and introduced directly into the biological system under study. This is similar to the first demonstrations that nitric oxide was the mysterious "EDRF" (endothelium derived relaxing factor) (Palmer et al., 1987; Ignarro et al., 1987).

**Acknowledgements**

The author thanks Drs. Loys Nunez, Edsel T. Bucovaz and Bob M. Moore for comments on the manuscript.

**References**

Bar-Nun, H. and Hartman, H. ((1978). Synthesis of organic compounds from carbon monoxide and water by UV photolysis. *Origins of Life*, **9**, 93-101.

Bass, G.E. and Crisan, D. (1973). Concerning irradiation-induced biological activity alterations of tetracycline. *Physiol. Chem. Phys*. **5**, 331-335.

Bass, G.E. (1975). The Comorosan effect: towards a perspective. *Int. J. Quantum Chem., Quantum Biol. Symp*. **2**, 321.

Bass, G.E., Sandru, D., Chenevey, J.E. and Bucovaz, E.T. (1976a). The Comorosan effect: single- and double-blind studies on the urea/urease system. *Physiol. Chem. Phys.* **8**, 253-258.

Bass, G.E. and Chenevey, J.E. (1976b). Irradiation induced rate enhancements for the LDH-pyruvate reaction. *Int. J. Quant. Chem. Quant. Biol. Symp*. **3**, 247-250.

Bass, G.E. and Chenevey, J.E. (1977). Substrate irradiation stimulation of the in vitro lactate-pyruvate interconversion reactions mediated by lactic dehydrogenase. *Physiol. Chem. Phys*. **9**, 555-562.

Boehning, D. and Snyder, S.H. (2002). Carbon monoxide and clocks. *Science*, **298**, 2339-2340.

Cirak, F., Cisternas, J.E., Cuitiño, A.M., Ertl, G., Holmes, P., Kevrekidis, I.G., Ortiz, M., Rotermund, H.H., Schunack, M. and Wolff, J. (2003). Oscillatory thermomechanical instability of an ultrathin catalyst. *Science*, **300**, 1932-1936.

Clyne, M.A.A. & Thrush, B.A. (1962). Mechanism of chemiluminescent combination reactions involving oxygen atoms. *Proc. R. Soc. Lond.* **A 269**, 404-418.

Comorosan, S. (1968). A quantum mechanical hypothesis for the mechanism of enzyme reactions. *Enzymol*. **35**: 117-130.

Comorosan, S., Sandru, S. and Alexandrescu, E. (1970a). Oscillatory behavior of enzymic reactions: a new phenomenology. *Enzymol.* **38**, 317-328.

Comorosan, S., Vieru, S., and Sandru, D. (1970b). A new approach to control mechanisms in tumor cells. *Europ. J. Can*. **6**, 393-400.

Comorosan, S, Vieru, S. and Sandru, D. (1970c). Evidence for a new biological effect of low-level irradiations. *Int.J.Radiat.Biol*. **17**, 105-115.

Comorosan, S (1970d). New mechanism for the control of cellular reactions: the biochemical flip-flop. *Nature* **227**, 64-65.

Comorosan, S., Vieru, S., Crisan, D., Sandru, D., and Murgoci, P. (1971a). A new metabolic control mechanism: I. Evidence for controlled biochemical networks in yeast cells. *Physiol. Chem. Phys*. **3**, 1-16.

Comorosan, S., Vieru, S., Crisan, D., Sandru, D., Murgoci, P., Alexandrescu, E. (1971b). A new metabolic control mechanism: II. Evidence for controlled biochemical networks in rabbit muscle tissue. *Physiol. Chem. Phys*. **3**, 103-115.

Comorosan, S., Murgoci, P., Sandru, D., and Cru, M. (1971c). A new metabolic control mechanism: III. Physico-chemical aspects of enzyme substrates perturbation. *Physiol. Chem. Phys*. **3**: 343-352.

Comorosan, S., Cru, M., Murgoci, P. and Vieru, S. (1972a). A new metabolic control mechanism: IV. Enzymes as measuring systems of substrates quantum properties. *Physiol. Chem. Phys.* **4**: 1-9.

Comorosan S., Cru, M. and Vieru, S. (1972b). The interaction between enzymic systems and irradiated substrates. *Enzymol*. **42**: 31-43.

Comorosan, S., Crisan, D., Alexandrescu, E. and Murgoci, P. (1972c). Investigation of rabbit aldolase isoenzymes with irradiated fructose-1,6-diphosphate. *Physiol. Chem. Phys.* **4**,:559-571.

Comorosan, S., Vieru, S. and Murgoci, P. (1972d). The effect of electromagnetic field on enzymic substrates. *Biochem. Biophys. Acta* **268**, 620-621.

Comorosan, S.,Murgoci, P., Vieru, S. and Crisan, D. (1973). The interaction of the electromagnetic field with organic molecules. *Studia Biophysica* **38**,169-175.

Comorosan, S. (1974). The measurement problem in biology. *Int. J. Quantum Chem.: Quantum Biol*. Symp. No. 1. 221-228.

Comorosan, S. (1975a). The measurement process in biological systems: a new phenomenolgy. *J. Theor. Biol.* **51**, 35-49.

Comorosan, S. (1975b). On a possible biological spectroscopy. *Bull. Math. Biol*. **37**, 419-425.

Comorosan, S. (1976). Biological obserables. In: *Progress in Theoretical Biology*, vol. 4 (Rosen, R. ed.) pp. 161-204, New York, Academic Press.

Comorosan, S., Hristea, M., Murgoci, P. (1980a). On a new symmetry in biological systems. *Bull. Math. Biol*. **42**, 107-117.

Comorosan, S., Tiron, V., Hristea, M., Cinca, S., Paslaru, L. and Popescu, V. (1980b). Interaction of water with irradiated molecules: a new physical aspect. *Physiol. Chem. Phys*. **12**, 497-504.

Comorosan, S., Jieanu, V., Morlova, I., Paslaru, L., Rozoveanu, P., Toroiman, E. and Vasilco, R. (1988). A novel interaction between light and matter: a review. *Physiol. Chem. Phys. Med NMR*, **20**: 319-328.

Dioum, E.M., Rutter, J., Tuckerman, J.R., Gonzalez, G., Gilles-Gonzalez, M.A., and McKnight, S.L. (2002) NPAS: a gas-responsive transcription factor. *Science*, **298**, 2385-2387.

Etzler, F.M. and Westbrook, D. (1986). Modulation of reaction kinetics via an apparently novel interaction of light and matter. *Physiol. Chem. Phys. Med. NMR* **19**, 271-274.

Goodwin, B.C. and Vieru, S. (1975). Low energy electromagnetic perturbation of an enzyme substrate. *Physiol. Chem. Phys*. **8**, 89-90.

Gray, P., Griffiths, J.F. & Scott, S.K. (1985). Oscillations, glow and ignition in carbon monoxide oxidation in an open system. I. Experimental studies of the ignition diagram and the effects of added hydrogen. *Proc. R. Soc. Lond*. **A 397**, 21-44.

Gray, P. & Scott, S.K. (1990). *Chemical Oscillations and Instabilities*, p. 410. Oxford: Clarendon Press.

Grisolia, S. (1975). communicated, $19^{th}$ Ann. Mtg., Biophysical Soc., Philadelphia.

Hobbs, P.V. (2000). *Introduction to Atmospheric Chemistry*, pp.24-25, Cambridge: Cambridge University Pres.

Hobbs, A.J. and Ignarro, L.J. (1996),.. Nitric oxide-cyclic GMP signal transduction system. In: *Methods in Enzymology*, **269**, 134-148, L. Packer, ed.

Hubbard, J.S., Hardy, J.P., and Horowitz, N.H. (1971). Photocatalytic Production of Organic Compounds from CO and $H_2O$ in a simulated Martian Atmosphere. *Proc. Natl. Acad. Sci. USA* **68**, 574-578.

Ignarro, L.J., Buga, G.M., Wood, K.S., Byrns, R.E. & Chaudhuri, G. (1987). Endothelium-derived relaxing factor produced and released from artery and vein is nitric oxide. *Proc. Natl. Acad. Sci. USA* **84**, 9265-9269.

Leffler, C.W., Balabanova, L., Sullivan, C.D., Wang, X., Fedinec, A.L. & Parfenova, H. (2003). Regulation of CO production in cerebral microvessels of newborn pigs. *Am. J. Physiol. Heart Circ. Physiol*, **285**, H292-H297.

Leguizamon, C.A., Cordero, J.M. & Zaretzky, A.N. (1987). A radiation induced periodic continuous effect on chemical kinetics detected by a photographic technique. *Physiol. Chem. Phys. Med. NMR*, **19**, 15-21.

Maines, M.D.(1996). Carbon monoxide and nitric oxide homology: differential modulation of heme oxygenase in brain and detection of protein activity. In: *Methods in Enzymology*, **268** (Parker, L., ed.) , pp. 473-488. New York: Academic Press.

Mukesh, D., Kenney, C.N., and Morton, W. (1983). Concentration oscillations of carbon monoxide, oxygen and 1-butene over a platinum supported catalyst. *Chem. Eng. Sci*., **38**, 69-77.

Nathan, C. (2002). Catalytic antibody bridges innate and adaptive immunity. *Science*, **298**, 2143-2144.

Oken, H.U. and Wicke, E. (1986). Statistical fluctuations of temperature and conversion at the catalytic CO oxidation in an adiabatic packed bed reactor. *Ber. Bunsenges. Phys. Chem.*, **90**, 976-81.

Palmer, R.M.J., Ferrige, A.G. & Moncada, S. (1987). Nitric oxide release accounts for the biological activity of endothelium-derived relaxing factor. *Nature* 327, 524-526.

Rashevsky, N. (1960). Contributions to relational biology. *Bull. Math. Biop.*, **22**, 73-84.

Rosen, R. (1961). On the role of chemical systems in the microphysical aspects of primary genetic mechanisms. *Bull. Math. Biop.*, **23**, 393-403.

Schwankner, R.J., Eiswirth, M., Moller, P., Wetzel, K., and Ertl, G. (1987). Kinetic oscillations in the catalytic CO oxidation on Pt(100): periodic perturbations. *J. Chem. Phys.*, **87**, 742-749.

Scott, R.P.and Watts, P. (1981). Kinetic considerations of mass transport in heterogeneous, gas-solid catalytic reactions. *J. Phys. E: Sci. Instrum.*, **14**, 1009-1013.

Sherman, R., Yee, W., Comorosan, S., Crisan, D. and Markovski, S. (1974). The effect of penicillin irradiation on bacterial growth and penicillin resistance. *Chemotherapy* **20**, 227-234.

Sherman, R.L., Siebert, S.T. and Yee, W.H. (1973). A note on the effect of electromagnetic field on enzymic substrates. *Physiol. Chem. Phys.*, **5**, 49-56.

Snyder, S.H. and Bredt (1992). Biological roles of nitric oxide. *Sci. Amer.*, May, 68-77.

Stamler, J.S., Singel, D.J. and Loscalzo, J. (1992). Biochemistry of nitric oxide and its redox-activated forms. *Science*, **258**, 1898-1902.

Wentworth, P., McDunn, J.E., Wentworth, A.D., Takeuchi, C., Nieva, J., Jones, T., Bautista, C., Ruedi, J.M., Gutierrez, A., Janda, K.D., Babior, B.M., Eschenmoser, A., and Lerner, R.A. (2002). Evidence for antibody-catalyzed ozone formation in bacterial killing and inflamation. *Science*, **298**, 2195-2199.

Winfree, A.T. (2002). On emerging coherence. *Science*, **298**, 2336-2337.

Wolin, M.S., Wood, K.S. & Ignarro, L.J. (1982). Gyanylate cyclase from bovine lung. *J. Biol. Chem.* **257**, 13312-13320.

Fig. 1. Typical enhancements induced in the lactic dehydrogenase reaction (pyruvate → lactate) with irradiated sodium pyruvate samples. From Comorosan, 1974.

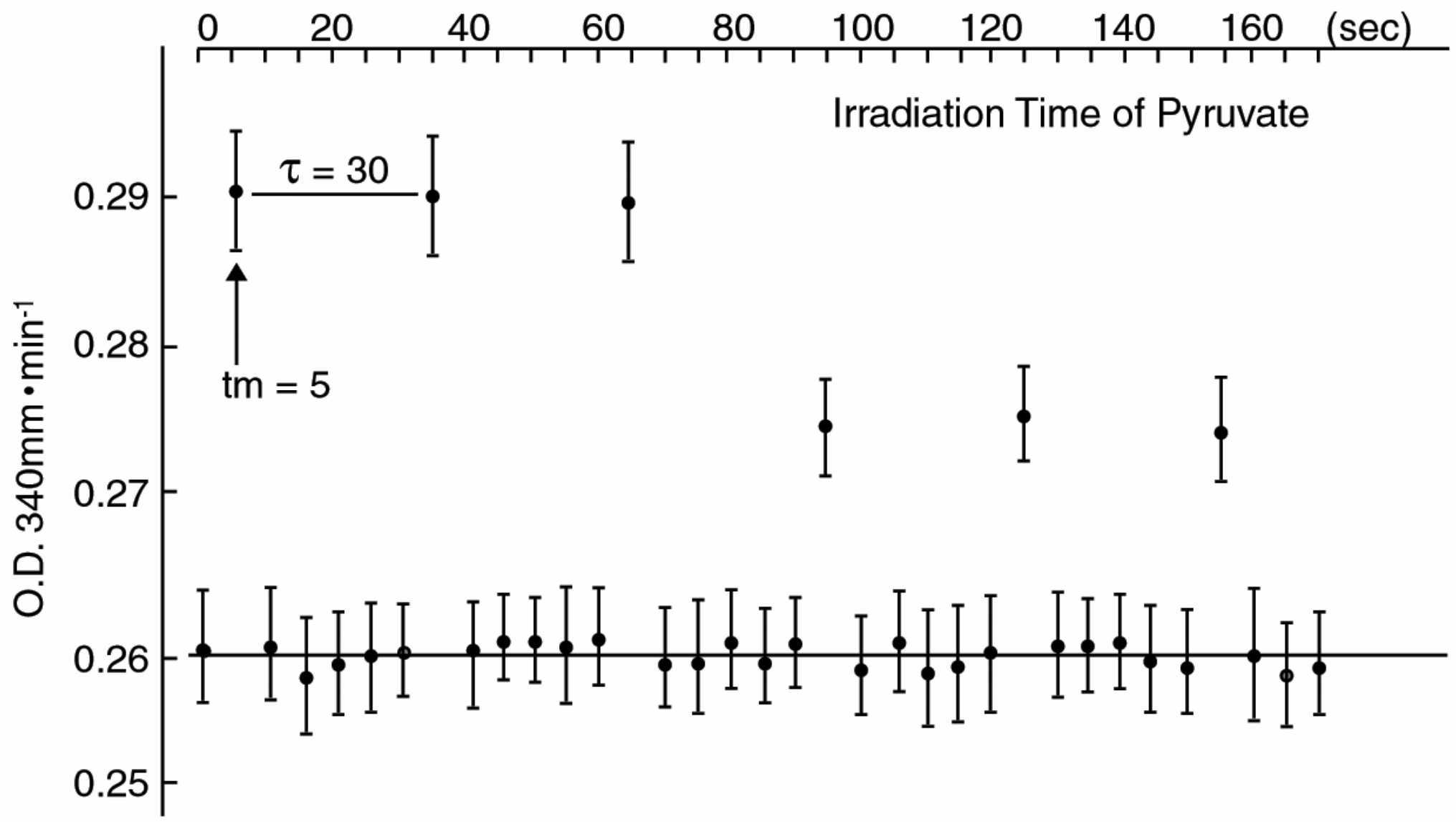

Fig. 2. Typical experimental records of oscillatory glow in the gas-phase CO + O reaction: (top) indefinitely sustained in a continuously fed stirred tank reactor; (bottom) a long, but finite, train for a closed system. Ordinate: light intensity, I, from chemiluminescence. Abscissa: reaction time in seconds. (From “Chemical Oscillations and Instabilities: Non-Linear Chemical Kinetics” by Gray, Peter & Scott, Stephen K., 1990, p. 427, by permission of Oxford University Press.)

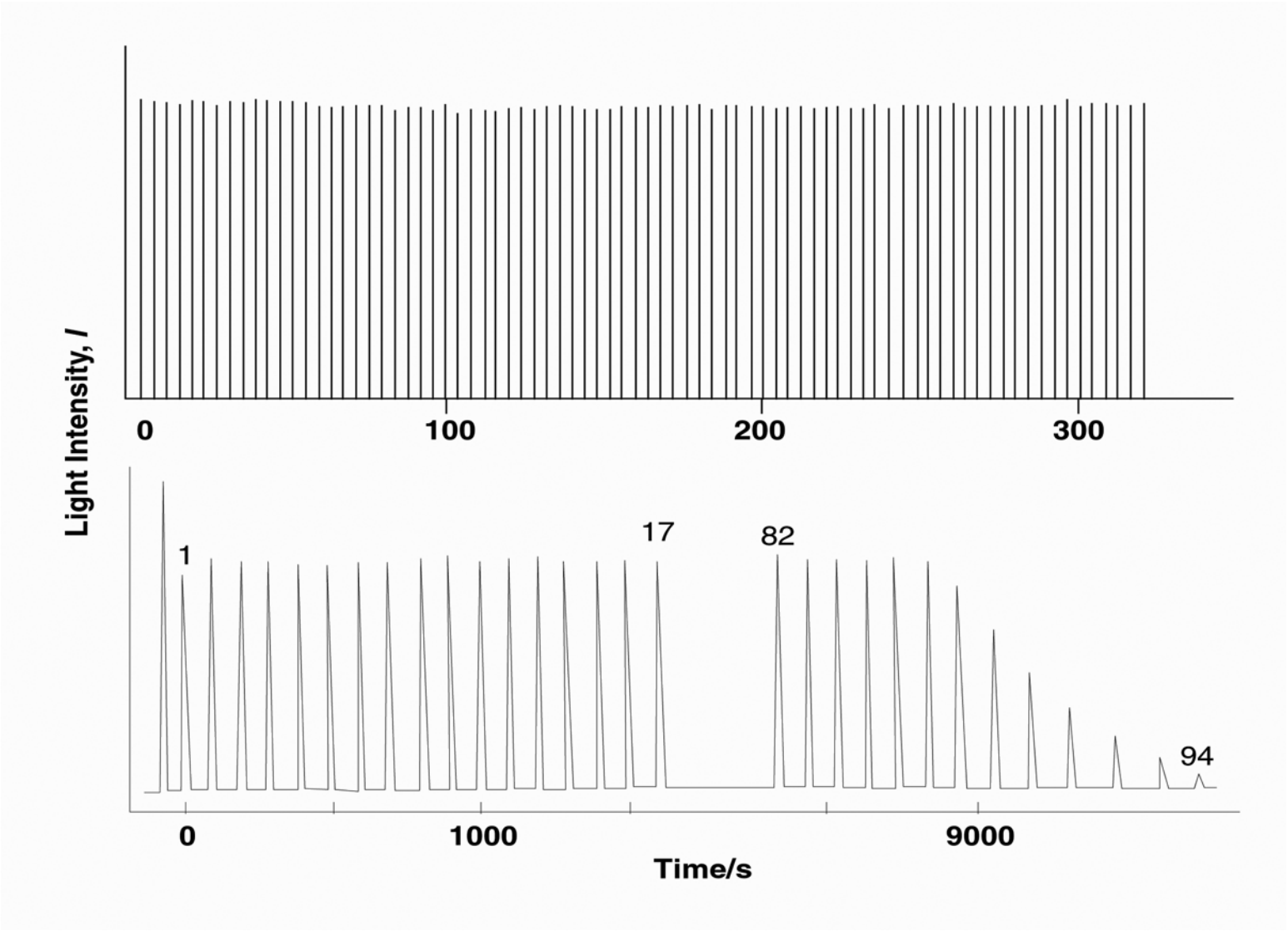

Fig. 3. Typical concentration, light emission, and temperature excess records for gas-phase oxidation of acetaldehyde in a continuously fed stirred tank reactor. Ordinate labels: Δc = concentration change for indicated species, I = light intensity for chemiluminescence, ΔT = temperature change due to exothermic reaction. Abscissa: reaction time in seconds. (From "Chemical Oscillations and Instabilities: Non-Linear Chemical Kinetics" by Gray, Peter & Scott, Stephen K., 1990, p. 432, by permission of Oxford University Press.)

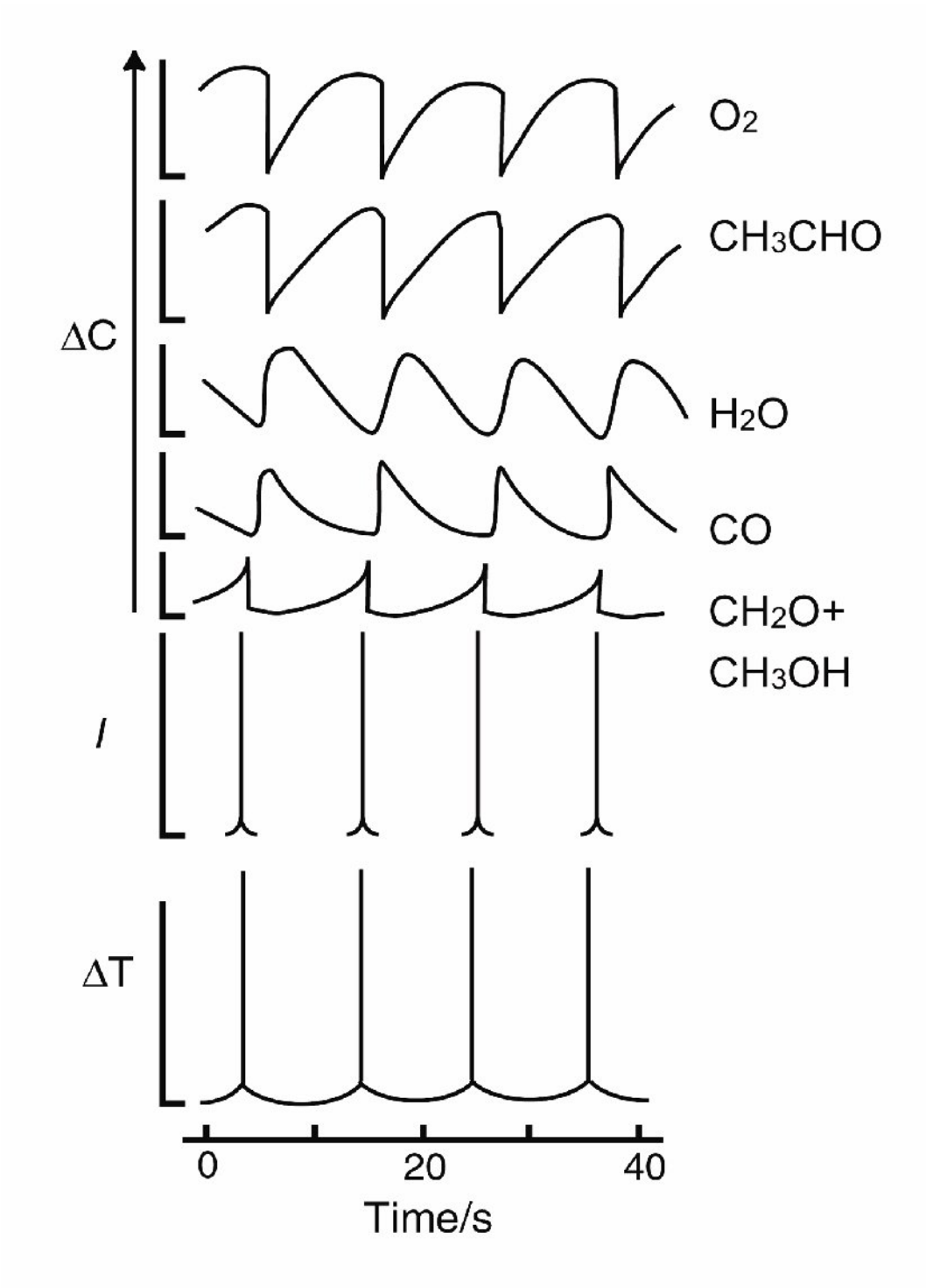

Fig. 4. Oscillatory oxidation of carbon monoxide on an ultrathin platinum catalyst. (A) The catalyst and substrate geometry. (B) deformation oscillations of the catalyst as time progresses, measured by integrating the gray-scale values within the square of 100x100 pixels shown in the third frame of (C), arb. u., arbitrary units. (C) Difuse light images from movie S1 of the catalyst surface deformations for a half-period at selected instants, as indicated by the four dots in (B). Images are 4.4 mm square. From Cirak et al., 2003.

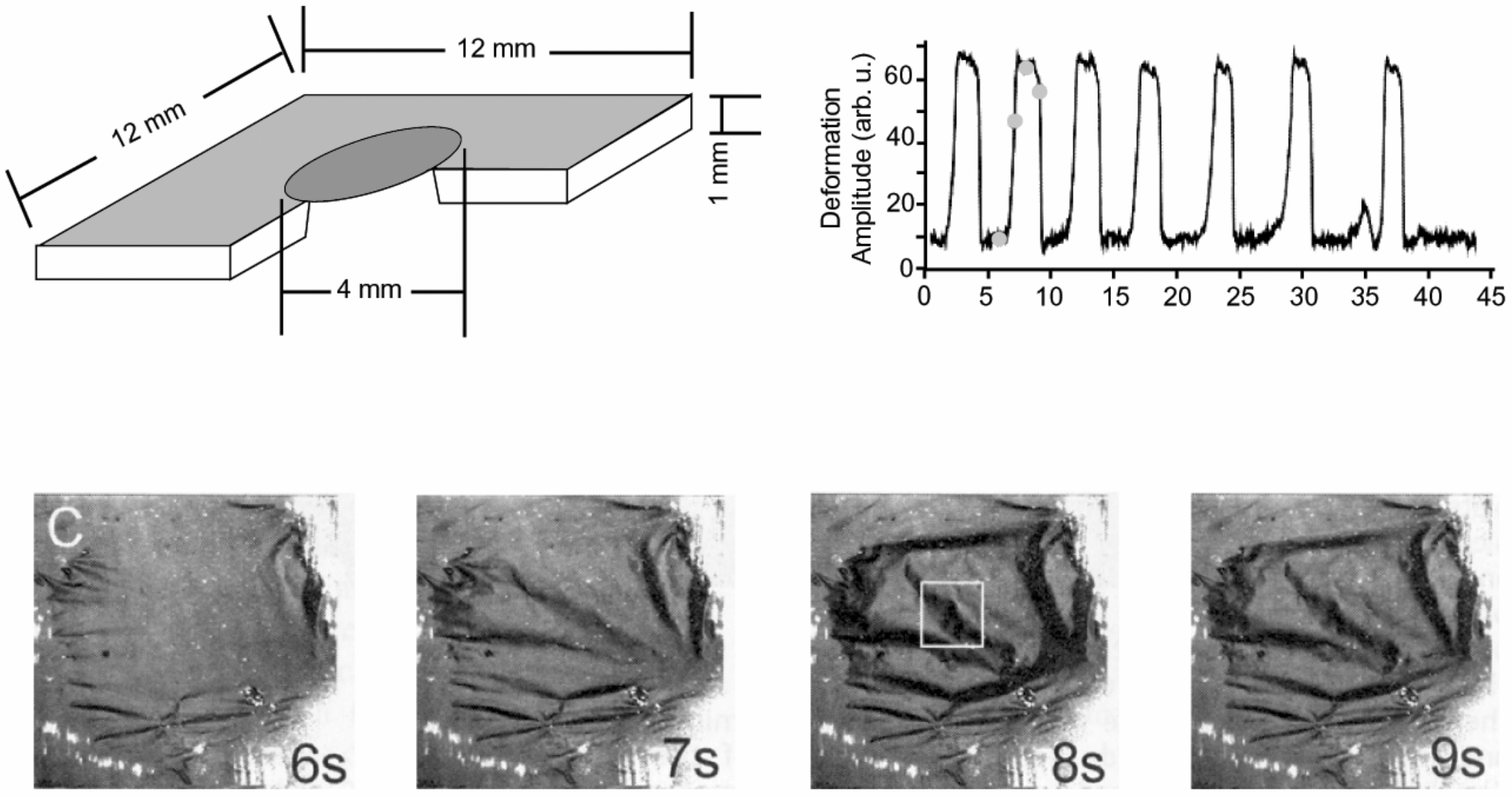

Table 1. Species, enzymes and substrates for which phenomenon has been observed.

| **Species** | **Enzymes** | **Irradiated Crystalline Substrates and Chemicals** |
|---|---|---|
| *Bacillus subtilis*<br>*Bacillus cereus*<br>*E. coli*<br>*Salmonella panama*[#]<br><br>*Saccharomyces cerevisiae* (yeast)<br><br>*Canavalia ensiformis* (jack bean)<br><br>rat<br>rabbit<br>pig<br>beef<br>human<br>chicken* | Aldolase<br>Citrate Synthase<br>Fumarase<br>Fructose-1,6-Diphosphatase<br>Glucose Dehydrogenase<br>Glucose-6-Phosphatase<br>Glucose-6-Phosphate Dehydrogenase<br>Glutamic Dehydrogenase<br>Glutamic-Oxalacetic Transaminase<br>Glutamic-Pyruvic Transaminase<br>Hexokinase<br>Invertase<br>Isocitrate Dehydrogenase<br>Lactate Dehydrogenase<br>Malate Dehydrogenase<br>Malic Enzyme<br>Penicillinase<br>Phosphoglucomutase<br>Phosphoglucose Isomerase<br>Phosphohexose Isomerase<br>Pyruvate Dehydrogenase<br>Succinate Dehydrogenase<br>Urease<br>Xanthine Oxidase | acetyl-Co-A<br>adenine<br>alanine<br>arginine<br>aspartate sodium<br>chloramphenicol hemisuccinate<br>cytidine<br>cytochrome C<br>fructose 1,6-diphosphate sodium<br>glucose<br>glucose 1-phosphate<br>glucose 6-phosphate<br>glutamate sodium<br>histidine<br>isocitrate sodium<br>α-ketoglutarate potassium<br>lactate lithium<br>malate sodium<br>6-mercaptopurine<br>mitomycin C<br>oxaloacetate<br>penicillin sodium<br>potassium chloride*<br>pyruvate sodium<br>silicon dioxide*<br>sodium bromide*<br>sodium chloride<br>succinate sodium<br>sucrose<br>tetracycline HCl<br>thymine<br>tryptophan<br>urea<br>xanthine sodium |

* this report
[#] unpublished, Bass et al.

Table 2. Sodium Chloride Irradiation Stimulation of the Pyruvate – Lactic Dehydrogenase (Rabbit Muscle) Reaction Rate

| **Run #** | $t^{h\upsilon}$ | **$\Delta A_{340}$ / 12-sec enzyme reaction**[a] | | | | $\Delta(\Delta A_{340})$ |
|---|---|---|---|---|---|---|
| | | **0 sec** | **$t^{h\upsilon}$ - 1 sec** | **$t^{h\upsilon}$ sec** | **$t^{h\upsilon}$ + 1 sec** | |
| 1 | 5 | 0.330 | 0.331 | 0.339 | 0.331 | 0.008 |
| 2 | 5 | 0.321 | 0.322 | 0.333 | 0.321 | 0.012 |
| 3 | 5 | 0.320 | 0.321 | 0.329 | 0.320 | 0.009 |
| | | | | | | |
| 4 | 35 | 0.328 | 0.329 | 0.338 | 0.329 | 0.009 |
| 5 | 35 | 0.315 | 0.315 | 0.324 | 0.315 | 0.009 |
| 6 | 35 | 0.316 | 0.315 | 0.324 | 0.315 | 0.009 |
| | | | | | | |
| 7 | 155 | 0.348 | 0.348 | 0.355 | 0.347 | 0.007 |
| 8 | 155 | 0.347 | 0.347 | 0.356 | 0.348 | 0.009 |
| 9 | 155 | 0.348 | 0.348 | 0.356 | 0.347 | 0.008 |
| | | | | | | |
| | | **0 sec** | **$t^{h\upsilon}$ sec** | **$t^{h\upsilon}$ sec** | **0 sec** | |
| 10 | 5 | 0.364 | 0.375 | 0.376 | 0.362 | 0.013 |
| 11 | 10 | 0.368 | 0.367 | 0.368 | 0.368 | -0.001 |
| 12 | 15 | 0.366 | 0.365 | 0.365 | 0.365 | 0.000 |
| 13 | 20 | 0.363 | 0.364 | 0.364 | 0.364 | +0.001 |
| 14 | 25 | 0.366 | 0.367 | 0.366 | 0.367 | 0.000 |
| 15 | 30 | 0.367 | 0.366 | 0.365 | 0.366 | -0.001 |
| 16 | 35 | 0.365 | 0.375 | 0.374 | 0.365 | 0.010 |
| | | | | | | |
| 17 | 995 | 0.351 | 0.359 | 0.360 | 0.352 | 0.008 |
| 18 | 995 | 0.350 | 0360 | 0.359 | 0.350 | 0.009 |
| 19 | 995 | 0.349 | 0.359 | 0.359 | 0.350 | 0.009 |
| | | | | | | |
| Summary: $t_m$ = 5 sec, $\tau$ = 30 sec | | | | | | |

[a]$\Delta A_{340}$ / 12-sec enzyme reaction = change in absorbance at 340nm in the first 12 seconds of the enzyme reaction.
**$t^{h\upsilon}$** = crystalline sodium chloride irradiation time, sec.
$\Delta(\Delta A_{340})$ = averaged ($\Delta A_{340}$ for $t^{h\upsilon}$ irradiated samples) – averaged ($\Delta A_{340}$ for non-$t^{h\upsilon}$ irradiated samples)

Table 3. Potassium Chloride Irradiation Stimulation of the Pyruvate – Lactic Dehydrogenase Reaction Rate

| **Run #** | $\mathbf{t^{h\upsilon}}$ | **$\Delta A_{340}$ / 12-sec enzyme reaction**[a] | | | | **$\Delta(\Delta A_{340})$** |
|---|---|---|---|---|---|---|
| | | **0 sec** | **$t^{h\upsilon}$ sec** | **$t^{h\upsilon}$ sec** | **0 sec** | |
| 1 | 5 | 0.355 | 0.364 | 0.365 | 0.356 | 0.009 |
| 2 | 5 | 0.354 | 0.364 | 0.363 | 0.354 | 0.009 |
| 3 | 5 | 0.355 | 0.363 | 0.363 | 0.355 | 0.008 |
| | | | | | | |
| 4 | 35 | 0.351 | 0.360 | 0.361 | 0.352 | 0.009 |
| 5 | 35 | 0.350 | 0.358 | 0.359 | 0.350 | 0.009 |
| 6 | 35 | 0.351 | 0.359 | 0.358 | 0.350 | 0.007 |
| | | | | | | |
| Summary: $t_m$ = 5 sec, $\tau$ = 30 sec | | | | | | |

[a]$\Delta A_{340}$ / 12-sec enzyme reaction = change in absorbance at 340nm in the first 12 seconds of the enzyme reaction.
$\mathbf{t^{h\upsilon}}$ = crystalline potassium chloride irradiation time, sec.
$\Delta(\Delta A_{340})$ = averaged ($\Delta A_{340}$ for $t^{h\upsilon}$ irradiated samples) – averaged ($\Delta A_{340}$ for non-$t^{h\upsilon}$ irradiated samples)

Table 4. Sodium Bromide Irradiation Stimulation of the Pyruvate – Lactic Dehydrogenase Reaction Rate

| **Run #** | $\mathbf{t^{h\upsilon}}$ | **$\Delta A_{340}$ / 12-sec enzyme reaction**[a] | | | | **$\Delta(\Delta A_{340})$** |
|---|---|---|---|---|---|---|
| | | **0 sec** | **$t^{h\upsilon}$ sec** | **$t^{h\upsilon}$ sec** | **0 sec** | |
| 1 | 5 | 0.344 | 0.352 | 0.353 | 0.345 | 0.008 |
| 2 | 5 | 0.343 | 0.353 | 0.352 | 0.344 | 0.009 |
| 3 | 5 | 0.342 | 0.351 | 0.350 | 0.343 | 0.008 |
| | | | | | | |
| 4 | 35 | 0.341 | 0.349 | 0.349 | 0.340 | 0.009 |
| 5 | 35 | 0.342 | 0.350 | 0.350 | 0.342 | 0.008 |
| 6 | 35 | 0.340 | 0.348 | 0.349 | 0.340 | 0.008 |
| | | | | | | |
| Summary: $t_m$ = 5 sec, $\tau$ = 30 sec | | | | | | |

[a]$\Delta A_{340}$ / 12-sec enzyme reaction = change in absorbance at 340nm in the first 12 seconds of the enzyme reaction.
$\mathbf{t^{h\upsilon}}$ = crystalline sodium bromide irradiation time, sec.
$\Delta(\Delta A_{340})$ = averaged ($\Delta A_{340}$ for $t^{h\upsilon}$ irradiated samples) – averaged ($\Delta A_{340}$ for non-$t^{h\upsilon}$ irradiated samples)

Table 5. Diatomaceous Earth Irradiation Stimulation of the Pyruvate – Lactic Dehydrogenase Reaction Rate: Unfiltered Suspension Added to Reaction Mixture.

| **Run #** | $\mathbf{t^{h\upsilon}}$ | **ΔA340 / 12-sec enzyme reaction**[a] | | | | **Δ(Δ A340)** |
|---|---|---|---|---|---|---|
| | | **0 sec** | $\mathbf{t^{h\upsilon}}$ **sec** | $\mathbf{t^{h\upsilon}}$ **sec** | **0 sec** | |
| 1 | 5 | 0.291 | 0.300 | 0.299 | 0.292 | 0.008 |
| 2 | 5 | 0.292 | 0.301 | 0.301 | 0.292 | 0.009 |
| 3 | 5 | 0.291 | 0.299 | 0.299 | 0.290 | 0.008 |
| | | | | | | |
| 4 | 35 | 0.290 | 0.299 | 0.299 | 0.291 | 0.009 |
| 5 | 35 | 0.291 | 0.300 | 0.298 | 0.290 | 0.008 |
| 6 | 35 | 0.289 | 0.298 | 0.299 | 0.290 | 0.009 |
| | | | | | | |
| Summary: $t_m$ = 5 sec, $\tau$ = 30 sec | | | | | | |

[a]$\Delta A_{340}$ / 12-sec enzyme reaction = change in absorbance at 340nm in the first 12 seconds of the enzyme reaction.
$\mathbf{t^{h\upsilon}}$ = crystalline diatomaceous earth irradiation time, sec.
$\Delta(\Delta A_{340})$ = averaged ($\Delta A_{340}$ for $t^{h\upsilon}$ irradiated samples) – averaged ($\Delta A_{340}$ for non-$t^{h\upsilon}$ irradiated samples)

Table 6. Diatomaceous Earth Irradiation Stimulation of the Pyruvate – Lactic Dehydrogenase Reaction Rate: Filtered[#] Solution Added to Reaction Mixture.

| **Run #** | **$t^{h\upsilon}$** | **$\Delta A_{340}$ / 12-sec enzyme reaction[a]** | | | | **$\Delta(\Delta A_{340})$** |
|---|---|---|---|---|---|---|
| | | **0 sec** | **$t^{h\upsilon}$ sec** | **$t^{h\upsilon}$ sec** | **0 sec** | |
| 1 | 5 | 0.287 | 0.295 | 0.296 | 0.288 | 0.008 |
| 2 | 5 | 0.289 | 0.296 | 0.296 | 0.289 | 0.007 |
| 3 | 5 | 0.288 | 0.296 | 0.295 | 0.288 | 0.008 |
| | | | | | | |
| 4 | 35 | 0.287 | 0.295 | 0.294 | 0.287 | 0.007 |
| 5 | 35 | 0.287 | 0.294 | 0.295 | 0.287 | 0.008 |
| 6 | 35 | 0.288 | 0.296 | 0.295 | 0.288 | 0.007 |
| | | | | | | |
| Summary: $t_m$ = 5 sec, $\tau$ = 30 sec | | | | | | |

[#] 0.22 micrometer Millipore filter.

[a]$\Delta A_{340}$ / 12-sec enzyme reaction = change in absorbance at 340nm in the first 12 seconds of the enzyme reaction.

**$t^{h\upsilon}$** crystalline diatomaceous earth irradiation time, sec.

$\Delta(\Delta A_{340})$ = averaged ($\Delta A_{340}$ for $t^{h\upsilon}$ irradiated samples) – averaged ($\Delta A_{340}$ for non-$t^{h\upsilon}$ irradiated samples)

Table 7. Comparison of Pyruvate Irradiation Activation for Chicken and Rabbit Muscle LDH

| Run | $t^{h\upsilon}$ | ΔA340 / 12-sec enzyme reaction [a] | | | | Δ(ΔA) | Δ(ΔA) |
|---|---|---|---|---|---|---|---|
| | | **0 sec** | **$t^{h\upsilon}$ sec** | **$t^{h\upsilon}$ sec** | **0 sec** | **Chicken** | **Rabbit** |
| 1 | 5 sec | R 0.245 | R 0.254 | C 0.349 | C 0.351 | -0.002 | +0.009 |
| 2 | | C 0.365 | C 0.365 | R 0.262 | R 0.252 | 0.000 | +0.010 |
| | | | | | | | |
| 3 | 10 sec | R 0.254 | R 0.254 | C 0.356 | C 0.357 | -0.001 | 0.000 |
| 4 | | C 0.359 | C 0.360 | R 0.256 | R 0.257 | +0.001 | -0.001 |
| | | | | | | | |
| 5 | 15 sec | R 0.257 | R 0.258 | C 0.375 | C 0.362 | +0.013 | +0.001 |
| 6 | | C 0.355 | C 0.366 | R 0.250 | R 0.250 | +0.011 | 0.000 |
| | | | | | | | |
| 7 | 20 sec | R 0.257 | R 0.257 | C 0.360 | C 0.359 | -0.001 | 0.000 |
| 8 | | C 0.364 | C 0.363 | R 0.255 | R 0.254 | -0.001 | +0.001 |
| | | | | | | | |
| 9 | 25 sec | R 0.255 | R 0.255 | C 0.359 | C 0.360 | -0.001 | 0.000 |
| 10 | | C 0.373 | C 0.372 | R 0.246 | R 0.248 | -0.001 | +0.002 |
| | | | | | | | |
| 11 | 30 sec | R 0.255 | R 0.253 | C 0.361 | C 0.363 | +0.002 | -0.002 |
| 12 | | C 0.362 | C 0.363 | R 0.253 | R 0.252 | +0.001 | +0.001 |
| | | | | | | | |
| 13 | 35 sec | R 0.275 | R 0.284 | C 0.367 | C 0.359 | +0.009 | +0.008 |
| 14 | | C 0.359 | C 0.368 | R 0.300 | R 0.291 | +0.009 | +0.009 |
| | | | | | | | |
| 15 | 55 sec | R 0.299 | R 0.299 | C 0.367 | C 0.358 | +0.009 | 0.000 |
| 16 | | C 0.358 | C 0.366 | R 0.292 | R 0.292 | +0.008 | 0.000 |
| | | | | | | | |
| Summary Chicken: $t_m$ = 15 sec, τ = 20 sec Rabbit: $t_m$ = 5 sec, τ = 30 sec | | | | | | | |

[a]$\Delta A_{340}$ / 12-sec enzyme reaction = change in absorbance at 340nm in the first 12 seconds of the enzyme reaction.

R = rabbit muscle LDH. C = chicken heart LDH

**$t^{h\upsilon}$** = crystalline sodium pyruvate irradiation time, sec.

$\Delta(\Delta A_{340})$ = ($\Delta A_{340}$ for $t^{h\upsilon}$ irradiated samples) – ($\Delta A_{340}$ for non-$t^{h\upsilon}$ irradiated samples) for rabbit muscle LDH or chicken heart LDH.

Table 8. Previously Reported $t_m$ and **τ** Parameters for Forward and Reverse Enzyme Reactions

| **Enzyme** | **Source** | **Irradiated Substrate** | **$t_m$ sec** | **τ sec** |
|---|---|---|---|---|
| Lactic dehydrogenase | Rabbit muscle | Pyruvate[a,b,c,d] | 5 | 30 |
| | | Lactate[a] | 20 | 30 |
| | | Lactate[b] | 15 | 30 |
| | Rat liver[e] | Pyruvate | 5 | 30 |
| | | Lactate | 15 | 30 |
| | Yeast[f] | Pyruvate | 5 | 20 |
| | | Lactate | 10 | 20 |
| | | | | |
| Malic dehydrogenase | Rabbit muscle[a] | Malate | 10 | 35 |
| | | Oxaloacetate | 15 | 35 |
| | Rat liver[e] | Malate | 10 | 35 |
| | | Oxaloacetate | 15 | 35 |
| | Yeast[f] | Malate | 5 | 20 |
| | | Oxaloacetate | 10 | 20 |
| | | | | |
| Malic enzyme | Rabbit muscle[a] | Pyruvate | 15 | 30 |
| | | Malate | 20 | 30 |
| | Yeast[f] | Pyruvate | 5 | 20 |
| | | Malate | 10 | 20 |
| | | | | |
| Gutamate pyruvate transaminase | Rabbit muscle[a] | Alanine | 15 | 30 |
| | | Pyruvate | 20 | 30 |
| | | | | |
| Phosphoglucomutase | Rabbit muscle[a] | Glucose-6-phosphate | 15 | 30 |
| | | Glucose-1-phosphate | 20 | 30 |

[a]Comorosan et al., 1971b
[b]Bass and Chenevey, 1977
[c]Sherman et al., 1973
[d]Goodwin and Vieru, 1975
[e]Comorosan et al., 1980
[f]Comorosan et al., 1971a

Table 9. Compilation of observed $t_m$ and $\tau$ combinations[a,b,c]

| | **$t_m$ , sec** | | | | | | | | | |
|---|---|---|---|---|---|---|---|---|---|---|
| **$\tau$, sec** | 5 | 10 | 15 | 20 | 25 | 30 | 35 | 40 | 45 | 50 |
| 5 | | | | | | | | | | |
| 10 | | | | | | | | | | |
| 15 | | | | | | | | | | |
| 20 | X | X | X | X | X | | | | | |
| 25 | X | | | | | | | | | |
| 30 | X | X | X | X | | | | | | |
| 35 | | X | X | | X | | | | X | |
| 40 | | | | | | | | | | |
| 45 | | | | | | | | | | |
| 50 | | | | | | | | | | |

[a] $t_m$ = minimum irradiation time producing reaction rate enhancement; $\tau$ = constant time period between successive reaction rate enhancing irradiation times.
[b]X indicates that the $t_m$, $\tau$ pair has been reported.
[c]All values taken from Comorosan et al., 1980b and this report.

**Appendix**

Materials and Methods

Identical 3 mg samples of sodium pyruvate (Sigma Chemical Co.) were prepared by pipetting 0.25 ml of a 12 mg/ml aqueous solution into identical small containers (Falcon 3001, 35x10 mm tissue culture dishes) which were placed in a vacuum desiccator at room temperature until dry (at least 6 h). These samples were irradiated for specific times with a narrow band selected from the output of a high pressure mercury lamp (General Electric H 100 A4/T, $\lambda_{max}$ = 546.1 nm, bandwidth = 8.8 nm, Detric Optics 2-cavity bandpass filter). The distance of the lamp above the sample (approximately 19 cm) had been adjusted to produce an intensity of 400 – 500 footcandles (Panlux Electronic Footcandle Meter) at the sample. The irradiated (and non-irradiated control) samples were dissolved in one ml of distilled water (3 min allowed for dissolution) and 0.1 ml aliquots placed in 1 cm square cuvettes along with 2.9 ml of 0.1 mg/ml NADH (Sigma Chemical Co.) / 0.05 M pH 7.5 phosphate buffer. Reaction was initiated in this solution, in place in the spectrophotometer, by addition of 0.050 ml of approximately 40 units/ml LDH (Sigma Chemical Co., Rabbit Muscle Type II and Chicken Heart). Absorbance change at 340 nm for the first 12 sec of reaction, $\Delta A_{340}$, was determined from the continuous, approximately linear absorbance recording utilizing a Gilford 2400S, chart speed 0.1 min/in., 0.5 absorbance units full scale. These settings permitted absorbance changes to be recorded to three significant figures. All experiments were conducted at room temperature (approximately 23 C). In order to take into consideration any variations in room temperature and to offset possible fluctuations in enzyme and NADH activities over the course of a day's experiments (and from one enzyme solution batch to another) the sodium pyruvate samples were assayed in groups of four, at least one of which was always a non-irradiated control. In the following, each of these groups of four samples is referred to as a "Run." In every Run, assay of the sample(s) irradiated for a t* time was always preceded and followed by assay of samples which were not irradiated or which were irradiated for non-t* times (also considered "control" samples). This allows distinction between irradiation-induced activation and possible activity drift of one sort or another in the course of a Run. In a given Run, all samples were first irradiated for the times indicated, then all were dissolved and loaded into cuvettes as outlined above. Following addition of NADH/buffer solution to all cuvettes, the four samples were assayed enzymatically within a lapse time of 10 min. The rapid manual addition and mixing of the enzyme was accomplished in 1.5 to 2.0 sec. Lapse time from beginning of the crystalline pyruvate dissolution step to completion of the enzyme assays was always less than 30 min. Dissolution of the samples was usually initiated within 5 – 10 min. after the irradiation step, always within 30 min. The crystalline sodium pyruvate samples were always used within 36 h following their preparation.

**Special Characters**

τ tau (Greek)
υ nu (Greek)
ε epsilon (Greek)

→ arrow for chemical reactions